\documentstyle[prl,aps, epsfig]{revtex}
\begin{document}
%\draft
%\baselineskip 0.7cm
\input{psfig.sty}
\def \gam {\frac{ N_f N_cg^2_{\pi q\bar q}}{8\pi} }
\def \gamm {N_f N_cg^2_{\pi q\bar q}/(8\pi) }
\def \be {\begin{equation}}
\def \ba {\begin{eqnarray}}
\def \ee {\end{equation}}
\def \ea {\end{eqnarray}}
\def \gap {{\rm gap}}
\def \gapp {{\rm \overline{gap}}}
\def \gappp {{\rm \overline{\overline{gap}}}}
\def \im {{\rm Im}}
\def \re {{\rm Re}}
\def \Tr {{\rm Tr}}
\def \P {$0^{-+}$}
\def \S {$0^{++}$}
\def \uu {$u\bar u$}
\def \dd {$d\bar d$}
\def \ss {$s\bar s$}
\def \qq {$q\bar q$}
\def \qqq {$qqq$}
\def \si {$\sigma(500-600)$}
\def \lsm {L$\sigma $M}
\title{Predicting $D\to\sigma\pi$}
\author{R. Gatto$^{(a)}$, G. Nardulli$^{(b)}$,
A.D. Polosa$^{(c)}$ and N. A. T\"ornqvist$^{(c)}$}
\address{(a) D\'epartement de Physique
Th\'eorique, Universit\'e de Gen\`eve, 24 quai E.-Ansermet,
CH-1211 Gen\`eve 4, Switzerland \\ (b) Dipartimento di Fisica,
Universit\`a di Bari and INFN Bari, via Amendola 173, I-70126
Bari, Italy \\ (c) Physics Department, POB 9, FIN--00014,
University of Helsinki, Finland}
\date{September, 2000} \maketitle
\begin{abstract}
We examine the $D\to\sigma\pi$ amplitude through a constituent
quark-meson model, incorporating heavy quark and chiral
symmetries, finding a good agreement with the recent E791 data
analysis of $D\to 3\pi$ via $\sigma$.
\\
\vskip 0.05cm \noindent Pacs numbers: 13.25.Ft, 12.39.Hg,
14.40.Cs\\ \noindent BARI-TH/00-393 \\  UGVA-DPT/2000-07-1086 \\
%HIP-2000-xxx/TH \vskip 0.90cm
\end{abstract}
%]

The light $\sigma$ resonance has accumulated considerable
theoretical and experimental interest after it was reintroduced as
a very broad resonance into the 1996 edition of the Reviews of
Particle Physics \cite{pdg}.  Recently a conference\cite{kyoto}
(at the Yukawa Institute for Theoretical Physics) was entirely
devoted to this controversial resonance. The broad $\sigma$  has
been difficult to disentangle from the available data, because the
analyses require sophisticated theoretical models, which apart
from unitarity, analyticity and coupled channels, also involve
constrains from chiral and flavour symmetries.

A direct experimental evidence seems to emerge from the
$D^+\to\sigma\pi^+\to 3\pi$ decay channel observed by the E791
collaboration \cite{E791}, where the $\sigma$ is seen as a clear
dominant peak covering $46\%$ of the $3\pi$ Dalitz plot. The
reason why it is so prominent in this reaction, is that the
background is small since S-waves dominate all subchannels, and
the Adler zero (which complicate the analysis in $\pi\pi\to\pi\pi$
since there it suppresses the low energy tail of the $\sigma$
signal) is absent in the  production of $\sigma$ in $D\to
\sigma\pi$. In this letter we shall adopt a
Constituent-Quark-Meson model, the CQM model \cite{rass}, to
calculate the amplitude for the process $D\to\sigma\pi$ and
compare this prediction with E791 data analysis. CQM is an
effective model that enables to calculate heavy meson decay
amplitudes through diagrams where  heavy mesons are attached at
the ends of loops containing heavy and light quark internal lines.
Essentially it is based on a Nambu-Jona-Lasinio effective
Lagrangian whose bosonization is responsible for effective
vertices (heavy meson)-(heavy quark)-(light quark). The model is
relativistic and incorporates the heavy quark symmetries and the
chiral $SU_2$ symmetry for the light quark sector of the
Lagrangian.

In the following we will make use of the heavy meson field
notation with $H$ and $S$ representing respectively the heavy
meson degenerate  $J^P$ doublets $(0^-,1^-)$ and $(0^+,1^+)$
predicted by Heavy-Quark-Effective-Theory (HQET) \cite{manowise}.
For our purposes $H$ and $S$ will represent the charmed mesons $D$
and $D(1^+)$. The physical characteristics of the latter have been
experimentally observed by the CLEO collaboration \cite{zoeller}.

We will focus on the {\it polar} and {\it direct} contributions
(shown in Figs. 1,2 and 3 respectively) to the $D\to\sigma$
semileptonic amplitude $\langle \sigma|A^\mu|D\rangle$. These
contributions have been extensively discussed also in the analysis
of the $B\to \pi$ semileptonic decay \cite{bipi}. We first
consider the polar contribution (Figs. 1,2).  This reduces to a
loop diagram in the CQM approach. According to the rules described
in \cite{rass}, the loop shown in Fig. 2, describing the amplitude
$\langle S(1^+)\sigma|H \rangle$, is computed through the
following integral:
\begin{equation}
(-1)i^3 (-i)^3 \sqrt{Z_H Z_S m_H m_S} \frac{N_c}{16\pi^4}\int^{\rm
reg} d^4l \frac{{\rm Tr}\left[(\gamma\cdot l+m)g_{\sigma qq
}(\gamma\cdot (l+q_\sigma)+m)\epsilon^\mu\gamma_\mu \gamma_5
\frac{1+\gamma \cdot v}{2}
(-\gamma_5)\right]}{(l^2-m^2)((l+q_\sigma)^2-m^2)(v\cdot
l+\Delta_H)}, \label{eq:loop}
\end{equation}
where $\epsilon^\mu$ is the polarization of the $S(1^+)$ state,
$q^\mu_\sigma=m_\sigma v^{\prime \mu}$, $\Delta_H=M_H-m_Q$, being
$M_H$ the mass of the incoming heavy meson (see Fig. 1,2), $v$ its
four-velocity  and $m_Q$ the mass of the constituent heavy quark
there contained ($m_Q=m_c$ in our case). As for the constants,
$(-1)$ comes from the fermion loop, $i^3$ from the three
propagators and $(-i)^3$ from the three vertices (the vertices
$HQq$ and $SQq$ are discussed in \cite{rass}, while the vertex
$qq\sigma$, brings the third factor of  $(-i)$ \cite{ebertbos})

After a continuation of the light propagator in the Euclidean
domain, the regularization prescription adopted for the
computation of the loop integral is the following:
\begin{equation}
\label{eq:regint} \int d^4 l_E \frac{1}{l^2_E + m^2}\to \int d^4
l_E\int_{1/\Lambda^2}^{1/\mu^2} ds e^{-s(l_E^2+m^2)}.
\end{equation}

The infrared and ultraviolet cutoffs $\mu$ and $\Lambda$ are
respectively $\mu=300$ MeV and $\Lambda=1.25$ GeV. The mass $m$ is
the constituent mass of the light quark as obtained by a NJL
gap-equation; its value is $m=300$ MeV. A discussion about the
choice of these values can be found in \cite{rass}. What should be
remarked here is that once fixed $\Lambda$ and $\mu$, the light
constituent mass $m$ is determined. Varying the cutoffs requires
also a recomputation of $m$. For $m$ values close to $300$ MeV,
infrared and ultraviolet cutoffs can range only in a narrow spread
of values.

The renormalization constants appearing in (\ref{eq:loop}) are:
\begin{eqnarray}
Z_H^{-1} &=& (\Delta_H+m) {\frac{\partial I_3(\Delta_H)}{\partial
\Delta_H}}
 +I_3(\Delta_H)\\
Z_S^{-1} &=& (\Delta_S-m) {\frac{\partial I_3(\Delta_S)}{\partial
\Delta_S}} +I_3(\Delta_S),
\end{eqnarray}
where:
\begin{eqnarray}
I_3(\Delta) &=& - \frac{iN_c}{16\pi^4} \int^{\mathrm {reg}}
\frac{d^4l}{(l^2-m^2)(v\cdot l + \Delta + i\epsilon)}\nonumber \\
&=&{N_c \over {16\,{{\pi }^{{3/2}}}}}
\int_{1/{{\Lambda}^2}}^{1/{{\mu }^2}} {ds \over {s^{3/2}}} \; e^{-
s( {m^2} - {{\Delta }^2} ) }\; \left( 1 + {\mathrm {erf}}
(\Delta\sqrt{s}) \right),
\end{eqnarray}
and $\Delta_S$ is defined in the same way as $\Delta_H$, i.e.,
$\Delta_S=M_S-m_Q$. $\Delta_H$ is the main free parameter of the
model. For it we choose three reasonable values
$\Delta_H=0.3,0.4,0.5$ GeV. All other quantities, including
$\Delta_S$, vary accordingly. The number of colours is $N_c=3$.
The coupling $g_{\sigma qq}$ of the $\sigma$ meson field to the
light quark fields emerges through the bosonization of a NJL
Lagrangian density leading to a linear $\sigma-$model of composite
fields, as discussed in \cite{ebertbos}:
\begin{equation}
g_{\sigma qq}=\frac{1}{2\sqrt{I_2}},
\end{equation}
where:
\begin{equation}
I_2=-\frac{iN_c}{16\pi^4} \int^{\mathrm {reg}} \frac{d^4l}{(l^2 -
m^2)^2}= \frac{N_c}{16\pi^2} \Gamma\left(0,\frac{m^2}{\Lambda^2},
\frac{m^2}{\mu^2}\right).
\end{equation}
Numerically:
\begin{equation}
g_{\sigma qq}=2.49.
\end{equation}

Defining the coupling:
\begin{equation}
\langle D(1^+)(p^\prime)\sigma (q_\sigma)
|D(p)\rangle=-iG_{DD(1^+)\sigma}\epsilon\cdot q_\sigma,
\label{eq:postul}
\end{equation}
the computation of the loop integral (\ref{eq:loop}) gives,
comparing (\ref{eq:loop}) and (\ref{eq:postul}):
% and using the
%general notation with $H$ and $S$ for $D$ and $D(1^+)$
\begin{equation}
G_{HS\sigma}=2g_{\sigma qq}C\sqrt{Z_H Z_S m_H m_S},
%\left[\frac{A_1}{m_H}+A_2\right],
\end{equation}
where $C$ is:
\begin{eqnarray}
%A_1 &=& I_3(\Delta_S)+2(m + \Delta_S-\Delta_H
%)\Omega_1-m_\sigma\Omega_2-2m^2 Z \\
C &=& \frac{1}{2
m_\sigma}\left(I_3\left(\frac{m_\sigma}{2}\right)-I_3\left(-\frac{m_\sigma}{2}\right)\right)
+(m+\Delta_H) Z - \frac{(\Delta_S-\Delta_H)-2m}{m_\sigma}\Omega_2,
\end{eqnarray}
and:
\begin{eqnarray}
Z &=&  \frac{iN_c}{16\pi^4} \int^{\mathrm {reg}}
\frac{d^4l}{(l^2-m^2)[(l+q)^2-m^2](v\cdot l + \Delta_1 +
i\epsilon)}\nonumber \\ &=&\frac{I_5(\Delta_1,
x/2,\omega)-I_5(\Delta_2,- x/2,\omega)}{2 x}.
\end{eqnarray}
In the case at hand, $\Delta_1=\Delta_H$, $\Delta_2=\Delta_S$,
$x=m_\sigma$ and:
\begin{equation} \omega=v\cdot v^{\prime}=\frac{v\cdot
q_\sigma}{m_\sigma}= \frac{v\cdot k-v\cdot
k^{\prime}}{m_\sigma}=\frac{\Delta_H-\Delta_S}{m_\sigma},
\end{equation}
where $k$ and $k^\prime$ are the residual momenta related to the
$H$ and $S$ fields as in Fig 2. The $I_5$ integral is computed
defining:
\begin{equation}
\eta(x,\Delta_1,\Delta_2,\omega)={{{\Delta_1}\,\left( 1 - x
\right) + {\Delta_2}\,x}\over {{\sqrt{1 + 2\,\left(\omega -1
\right) \,x + 2\,\left(1-\omega\right) \,{x^2}}}}}.
\end{equation}
The explicit expression is:
\begin{eqnarray}
I_5(\Delta_1,\Delta_2,\omega) &= & \frac{iN_c}{16\pi^4}
\int^{\mathrm {reg}} \frac{d^4l}{(l^2-m^2)(v\cdot l + \Delta_1 +
i\epsilon ) (v'\cdot l + \Delta_2 + i\epsilon )} \nonumber \\
 & = & \int_{0}^{1} dx \frac{1}{1+2x^2 (1-\omega)+2x
(\omega-1)}\times\nonumber\\ &&\Big[
\frac{6}{16\pi^{3/2}}\int_{1/\Lambda^2}^{1/\mu^2} ds~\eta \;
e^{-s(m^2-\eta^2)} \; s^{-1/2}\; (1+ {\mathrm {erf}}
(\eta\sqrt{s})) +\nonumber\\
&&\frac{6}{16\pi^2}\int_{1/\Lambda^2}^{1/\mu^2} ds \;
e^{-s(m^2-2\eta^2)}\; s^{-1}\Big].
\end{eqnarray}
Moreover:
\begin{equation}
\Omega_2=\frac{
-I_3(\Delta_1)+I_3(\Delta_2)-\omega[I_3(-x/2)-I_3(x/2)]} {2 x
(1-\omega^2)} - \frac{[x/2- \Delta_1\omega ]Z}{1-\omega^2},
\end{equation}
where again $\Delta_1=\Delta_H$, $\Delta_2=\Delta_S$ and
$x=m_\sigma$.

Let's write the weak current matrix element for the semileptonic
transition amplitude $D\to \sigma$:
\begin{eqnarray}
\langle \sigma(q_\sigma)|A^{\mu}(q)|D(p)\rangle &=& \left[
(p+q_\sigma)^{\mu}+\frac{m_{\sigma}^2-m_D^2}{q^2}q^\mu \right]\;
F_1(q^2)\nonumber \\ &-& \left[
\frac{m_{\sigma}^2-m_D^2}{q^2}q^{\mu} \right] \; F_0(q^2),
\label{eq:effezero}
\end{eqnarray}
with $F_1(0)=F_0(0)$. Defining:
\begin{equation}
\langle {\rm VAC}|A^\mu|D(1^+)\rangle = \hat{F}^+
\sqrt{m_{D(1^+)}}\epsilon^\mu,
\end{equation}
we have:
\begin{equation}
F_1(q^2)=\frac{1}{2}\times \frac{\sqrt{m_{D(1^+)}}\hat{F}^+
G_{DD(1^+)\sigma}}{m_{D(1^+)}^2-q^2}.
\end{equation}
This is equivalent to assuming a polar model for the form factor
$F_1(q^2)$ with $D(1^+)$ taken as the intermediate virtual state
(see Figs. 1,2). The polar form factors are obviously more
reliable near the pole, where $q^2\simeq m_S^2$, than in the small
$q^2$ range. Anyway we assume the polar behavior valid for the
whole $q^2$ range. For our purposes the $\sigma$ meson is a
$J^{PC}=0^{++}$ isoscalar with a quark content
$(\bar{u}u+\bar{d}d)$.

Using the values $m_\sigma=478$ MeV \cite{E791}, $m_S=2.461$ GeV
\cite{zoeller}, $m_H=1.869$ GeV \cite{pdg} and $\hat{F}^+$ given
by the CQM model as a function of $\Delta_H$ \cite{rass} we have,
varying $\Delta_H$ (and consequently $\Delta_S$, see \cite{rass})
in the range of values $0.3,0.4,0.5$ GeV:
\begin{equation}
F_1^{\rm pol}(0)=0.30 \pm 0.04. \label{eq:risultato}
\end{equation}

The same calculation for $F_0(q^2)$ implies to consider the $D$
meson in the dispersion relation. Assuming that we can extrapolate
the polar behaviour to $q^2=0$ leads to:
%Since it is not clear if
%$F_0^{(D\sigma)}(q^2)$ has a polar dependence, whereas this is
%well understood for the $F_0^{(D\pi)}(q^2)$
%\cite{bipi,reportc}(and references therein), as a further check we
%also compute the $F_0^{\rm pol}(0)$ for the $D\to \sigma$
%transition extracting it from a polar diagram, such that in Fig.
%2, where the intermediate polar state is the $D(0^-)$. This leads
%to:
\begin{equation}
F_0^{\rm pol}(0)=\frac{\hat{F} G_{DD\sigma}}{2m_D^{5/2}},
\end{equation}
where $\hat{F}$ is defined by:
\begin{equation}
\langle {\rm VAC}|A^\mu|D(p)\rangle = i p^\mu
\frac{\hat{F}}{\sqrt{m_D}},
\end{equation}
and is computed in CQM \cite{rass}; $G_{DD\sigma}$ is given by:
\begin{equation}
G_{DD\sigma}=2g_{\sigma qq} Z_H m_H (2m \Omega_1 -m^2 Z),
\end{equation}
with:
\begin{equation}
\Omega_1=\frac{ I_3(-x/2)-I_3(x/2)+\omega[I_3(\Delta_1)-
I_3(\Delta_2)]}{2 x (1-\omega^2)} - \frac{[\Delta_1-\omega x/2]Z}
{1-\omega^2},
\end{equation}
where $\Delta_1=\Delta_H=\Delta_2$ and $\omega=0$. The numerical
result is:
\begin{equation}
F_0^{\rm pol}(0)=0.22_{-0.01}^{+0.07}.
\end{equation}
%in reasonable agreement with the $F_1(0)=F_0(0)$ requirement
%(needed to avoid the $q^2=0$ singularity in (\ref{eq:effezero})).
%This for what concerns the polar contribution to the form factors
%(see Figs. 1,2). CQM predicts also a {\it direct} contribution,
%see Fig. 3,  where the weak current $A^\mu$ is attached directly
%to the loop, without an intermediate $S$ or $H$ state. This
%contribution is computed through a loop integral similar to that
%given in (\ref{eq:loop}) where $m_S, Z_S$ and $\epsilon^\mu$
%should not be included. Extracting the $F_1(q^2)$ component one
%finds:

Let us now consider the {\it direct} contribution of Fig. 3,
obtaining:
\begin{equation} F_1^{\rm dir}(q^2)=2\sqrt{Z_H m_H}g_{\sigma qq}\left(
\frac{c \Omega_1}{2 m_H}+\frac{c\Omega_2}{2 m_\sigma}-\frac{a}{2
m_H}-\frac{b}{2}\right),
\end{equation}
where:
\begin{eqnarray}
a &=& -I_3(\Delta)+2m^2 Z+m_\sigma (\omega\Omega_1 +
\Omega_2)\nonumber\\ b &=& \frac{1}{2
m_\sigma}\left[I_3\left(-\frac{m_\sigma}{2}\right)-I_3\left(\frac{m_\sigma}{2}
\right)\right]-(\Delta_H+m)Z \nonumber\\ c&=&(m_\sigma \omega
+2m), \nonumber
\end{eqnarray}
and we notice that $\omega=\frac{m_D^2+m_\sigma^2-q^2}{2 m_D
m_\sigma}$, $\Delta=\Delta_H-m_\sigma \omega$,
$\Delta_1=\Delta_H$, $\Delta_2=\Delta$, $x=m_\sigma$. Numerically:
\begin{equation}
F_1^{\rm dir}(q^2=0)=0.30\pm 0.02.
\end{equation}

The analogous result for $F_0^{\rm dir}(q^2)$ is:
\begin{equation}
F_0^{\rm dir}(q^2)=2\sqrt{Z_H m_H}g_{\sigma qq}\left[\left(\frac{c
\Omega_1-a}{2
m_H}\right)\left(1+\frac{q^2}{m_D^2-m_\sigma^2}\right) +
\left(\frac{c\Omega_2}{2
m_\sigma}-\frac{b}{2}\right)\left(1-\frac{q^2}{m_D^2-m_\sigma^2}\right)\right]
\end{equation}
with $F_0^{\rm dir}(q^2=0)=0.30\pm 0.02$. We conclude that the
CQM-model analysis gives:
\begin{equation}
F_0^{\rm pol}(0)+F_0^{\rm dir}(0)=0.52^{+0.09}_{-0.03},
\label{eq:risultatovr}
\end{equation}
We have not included in this analysis the uncertainty arising from
the extrapolation to $q^2=m_\pi^2\simeq 0$ of the result obtained
by the dispersion relation, strictly valid only for $q^2\simeq
m_D^2$. We can estimate it by considering that $F_1(0)=F_1^{\rm
pol}(0)+F_0^{\rm dir}(0)$  should be equal to $F_0(0)$. Our result
for $F_1(0)$ is:
\begin{equation}
F_1(0)=0.60 \pm 0.06,
\end{equation}
which agrees within errors with the number obtained in
(\ref{eq:risultatovr}). Our estimate is therefore:
\begin{equation}
 F_0(m_\pi^2) \simeq F_0(0)=0.57\pm 0.09.
\label{eq:finale}
\end{equation}
This result has to be compared with that obtained directly from
preliminary E791 data \cite{dibandscad}:
\begin{equation}
F_0(m_\pi^2)=0.79\pm 0.15,
\end{equation}
by means of the following expression for the $D\to \sigma\pi$
amplitude:
\begin{equation}
\langle \sigma \pi^+|H_{\rm
eff}|D^+\rangle=\frac{G_F}{\sqrt{2}}V^*_{cd}V_{ud}a_1
F_0(m_\pi^2)(m_D^2-m_\sigma^2)f_\pi,
\end{equation}
where $H_{\rm eff}$ is the effective Hamiltonian of Bauer, Stech
and Wirbel \cite{BSW}, with $a_1=1.10\pm 0.05$ fitted for $D$
decays while  the value for the amplitude $\langle \sigma
\pi^+|H_{\rm eff}|D^+\rangle$ is computed considering the
experimental evidence for
$\Gamma(D^+\to\sigma\pi^+\to\pi^+\pi^-\pi^+)=0.44\times
\Gamma(D^+\to\pi^+\pi^-\pi^+)$ and taking the strong coupling
constant $g_{\sigma\pi\pi}$ derived from the preliminary E791 fit
for the sigma width $\Gamma_\sigma=338\pm 48$ MeV
\cite{dibandscad}.

%CQM therefore predicts for the full three body decay width of $D$
%into three pions via $\sigma$ (discussed in \cite{dibandscad}) a
%value that is consistent with $0.50\times
%\Gamma(D^+\to\pi^+\pi^-\pi^+)$ (the E791 collaboration preprint
%\cite{E791} describes a fit in which the $\sigma$ amplitude
%produces the largest decay fraction, $46\%$ with a statistical
%error of $9\%$).

We are aware of the theoretical uncertainties of the present
calculation; in particular, $1/m_c$ corrections, that have been
neglected in the quark loop calculation and in the evaluation of
$\hat{F}^+$, may alter our result (\ref{eq:finale}). To estimate
these uncertainties we note that CQM can be applied to the
evaluation of the coupling $F_1^{(D\pi)}(0)$ for which
experimental data are also available. We observe that in the case
of the $D\to\pi$ semileptonic process, the polar form factor
$F_1^{(D\pi)}(0)$ can be obtained from $F_1^{(B\pi)}(0)$, computed
in \cite{bipi} by CQM, simply using the following scaling form
\cite{reportc}:
\begin{equation}
F_1^{(D\pi)}(0)=\sqrt{\frac{m_B}{m_D}}F_1^{(B\pi)}(0)=0.87\pm
0.02,
\end{equation}
(neglecting small QCD corrections) since the computation of
$F_1^{(B\pi)}(0)$ is more stable against $1/m_Q$ corrections. This
must be compared with the $F^{(D\pi)}_1(0)=0.78\pm 0.06$,
deducible from the PDG \cite{pdg}, indicating that $1/m_c$
corrections are not so strong to qualitatively compromise the
results. Our analysis does not throw light on the fundamental
nature of the $\sigma$ resonance, whose theoretical status remains
uncertain. Independently of the actual nature of the signal, it
shows, however, that its decay properties can be understood and
predicted in a well defined and reasonable model, the CQM model.
Numerically its weak coupling to the current and to the charmed
mesons is similar to that of the pseudoscalar bosons, a result
which we believe robust and independent on the details of the
actual model we have used in the present letter.

\acknowledgements

We would like to thank  Prof. A. Deandrea for discussion on the
subject. ADP acknowledges correspondence with Prof. C. Shakin and
Prof. M.K. Volkov. ADP and NAT acknowledge support  from EU-TMR
programme, contract CT98-0169.

%%%%%%%%%%%%%%%%%%%%%%%%%%%%%%%%%%%%%%%%%%%%%%%%%%%%%%%%%%%%%%%%%%
\newpage
%%%%%%%%%%%%%%%%%%%%%%%%%%%%%%%%%%%%%%%%%%%%%%%%%%%%%%%%%%%%%%%%%%
%===============================================================================
\begin{figure}[t!]
\begin{center}
\epsfig{bbllx=0.5cm,bblly=16cm,bburx=20cm,bbury=23cm,height=5truecm,
        figure=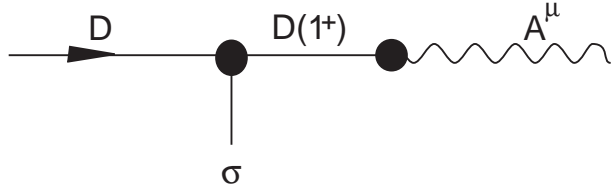}
\caption{\label{fig:fig1}
\footnotesize
          Diagram for the polar contribution to the $D\to\sigma$ semileptonic
          amplitude. }
\end{center}
\end{figure}
%===============================================================================
%===============================================================================
\begin{figure}[t!]
\begin{center}
\epsfig{bbllx=0.5cm,bblly=16cm,bburx=20cm,bbury=23cm,height=5truecm,
        figure=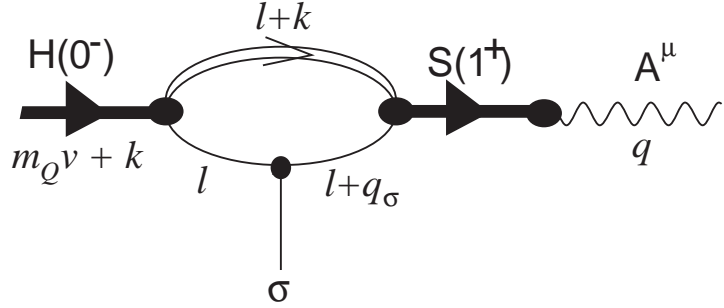}
\caption{\label{fig:fig2}
\footnotesize
         The CQM loop diagram for the process in Fig. 1. $H$ and $S$ are the
         heavy meson fields of Heavy-Quark-Effective-Theory,
         while $l$ is the momentum running in the loop. Here
         $H$ and $S$ coincide respectively with $D$ and $D(1^+)$. The residual momentum
         carried by the $S$ field is $k^\prime$ while its four velocity is still $v$
         since no external current is acting on the heavy quark line (the doubled
         one).\newline
         The weak current $A^\mu$ could be {\it directly} attached to the loop (i.e., without
         an intermediate $S$ state). This kind of contribution to the form factors,
         is shown in Fig. 3.}
\end{center}
\end{figure}
%===============================================================================
%===============================================================================
\begin{figure}[t!]
\begin{center}
\epsfig{bbllx=0.5cm,bblly=16cm,bburx=20cm,bbury=23cm,height=5truecm,
        figure=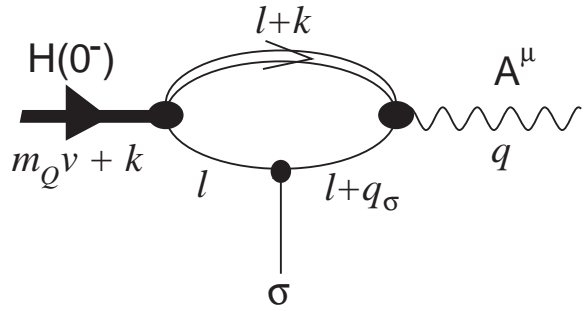}
\caption{\label{fig:fig3} \footnotesize
          Diagram for the direct contribution to the $D\to\sigma$ semileptonic
          amplitude. }
\end{center}
\end{figure}
%===============================================================================

\end{document}